\documentclass[twocolumn,showpacs,prl,amsmath,amssymb]{revtex4}
\usepackage{graphicx}
\usepackage{epsfig}
\usepackage{dcolumn}
\usepackage{bm}
\usepackage{epstopdf}
\usepackage{color}

\begin{document}
\title{Charge-induced spin polarization in non-magnetic organic molecule Alq$_{3}$}
\author{K. Tarafder}
\author{B. Sanyal}
\author{P. M. Oppeneer}
\affiliation{Department of Physics and Astronomy, Uppsala University, Box 516, SE-75120 Uppsala, Sweden}
\date{\today}
\begin{abstract}
Electrical injection in organic semiconductors is a key prerequisite for the realization of organic spintronics.
Using density-functional theory calculations we report the effect of electron transfer into the organic molecule Alq$_3$. 
Our first-principles simulations  show that electron injection spontaneously spin-polarizes non-magnetic Alq$_3$ with a magnetic moment linearly increasing with induced charge. An asymmetry of the Al--N bond lengths leads to an asymmetric distribution of injected charge over the molecule. The spin-polarization arises from a filling of dominantly the nitrogen $p_z$ orbitals in the molecule's LUMO together with ferromagnetic coupling of the spins on the quinoline rings.

\end{abstract}
\maketitle

Organic semiconductors (OSCs) are contemporary being considered as prospective materials for spintronics applications \cite{xiong04,naber07}.
As OSCs primarily consist of $\pi$-conjugated molecules composed of atoms with low atomic numbers, the spin-orbit and nuclear hyperfine interactions are weak. OSCs therefore exhibit typically very long spin relaxation times \cite{pramanik07}, a property favorable for spin transport. 
Moreover, owing to the availability of a large variety of organic semiconductors (OSCs) and the possibility to chemically tailor specific functionalities, there is a huge potential for the development of versatile, low-cost organic spintronic devices \cite{dediu09,pulizzi09}.

A key pre-requisite for the realization of organic spintronics is the realization of electrical injection in the organic layer. Recently, the injection of spin-polarized carriers from a ferromagnetic contact into an OSC has been demonstrated \cite{cinchetti09,drew09}.
Notwithstanding this affirmative observation, not all OSC materials are comparably well suited for organic spintronics.
In particular, the $\pi$-conjugated OSC  tris(8-hydroxy-quinoline) aluminum (Alq$_{3}$) has been discovered  to perform very good in organic spin-valves, providing one of the best giant magnetoresistance (GMR) values \cite{xiong04}.
Alq$_3$ is a relatively well-known material that is widely used as component in organic light-emitting diodes (OLEDs) \cite{tang87,vanslyke96}. However, in spite of this, it is not understood why precisely Alq$_3$ 
%performs so well in GMR devices.
gives rise to extremely long spin-relaxation times \cite{pramanik07}, to a high GMR \cite{xiong04,note}, and significant tunnel magnetoresistance (TMR) values \cite{santos07}.

Here we employ density functional theory (DFT) calculations to explore the influence of electron injection into Alq$_3$. Our first-principles calculations reveal that electron charging causes, unexpectedly, a spin-polarization of the organic molecule, which is non-magnetic in its neutral state. The origin of the appearing spin-polarization is analyzed in detail and shown to be connected to the small asymmetry of the $\pi$-conjugated rings, i.e., a feature that is closely related to the particular molecular geometry and molecular orbitals of Alq$_3$. 
The here-reported influence of electron injection in Alq$_3$ provides an essential ingredient
%n important paradigm (?) 
to a microscopic understanding of the complex spin-polarized transport in Alq$_3$.

%previous investigations 
Since Alq$_3$ is widely used as a material  in organic electro-luminescent devices \cite{tang87,vanslyke96}, a number of experimental as well as theoretical studies have been performed for this material. Using DFT calculations, photo-emission, and near-edge x-ray absorption fine structure (NEXAFS), Andreoni and co-workers \cite{curioni98} have characterized the structural and electronic properties of Alq$_{3}$ in both neutral and charged states, which provided a 
%complete
picture of the orbital structure of Alq$_{3}$. Johansson {\it et al.} \cite{johansson99} used a combination of x-ray and ultraviolet photo-emission spectroscopy (UPS) with DFT calculations to discuss the interaction of Alq$_{3}$ with Li and K. Recently, the geometrical and electronic structures of Alq$_{3}$ interacting with Mg and Al have also been investigated using DFT calculations \cite{meloni03, zhang02}. Baik {\it et al.} \cite{baik08} tried to make this material magnetic by Co doping. 
Zhan {\it et al.} \cite{zhan10} studied the interaction of Alq$_3$ with an Fe surface.
Using DFT, semi-empirical, and {\it ab initio} molecular orbital theories, Zhang {\it et al.} \cite{zhangcpl} studied the effect of electrical charging on the electronic structure of Alq$_{3}$. However, these studies have not considered any geometric structural change in the molecule due to charge injection, nor any magnetic effects. 

%\subsection{Computational Details}
We have performed first-principles DFT calculations using the full-potential Vienna {\it ab initio} simulation package (VASP) \cite{vasp} which uses pseudopotentials and the projector augmented wave approach. We used the Perdew-Burke-Ernzerhof generalized gradient approximation (GGA) for treating the exchange-correlation potential. A sufficiently high energy cutoff (550 eV) was used in each calculation to obtain accurate results. 
%Using iterative matrix diagonalization  with conjugate gradient method, the Kohn-Sham equations were solved self-consistently. 
In our simulations the forces on each of the atoms were calculated using the Hellmann-Feynman theorem, and were subsequently used to perform a conjugate gradient structural relaxation.
The structural optimizations were continued until the forces on the atoms converged to less than 
 1 meV/{\AA}. This optimization has been completely carried through for each Alq$_3$ molecule with an additional charge. There are two isomers of Alq$_{3}$, namely {\sl facial} and {\sl meridianal} with C$_3$ and C$_1$ point group symmetries, respectively. Of these two, the meridianal isomer is reported \cite{curioni98,halls01} to possess the lowest formation energy. This we have also confirmed in our calculations. In the following we consider therefore only meridianal Alq$_3$.

\begin{figure}[!tb]
\includegraphics[width=0.45\textwidth]{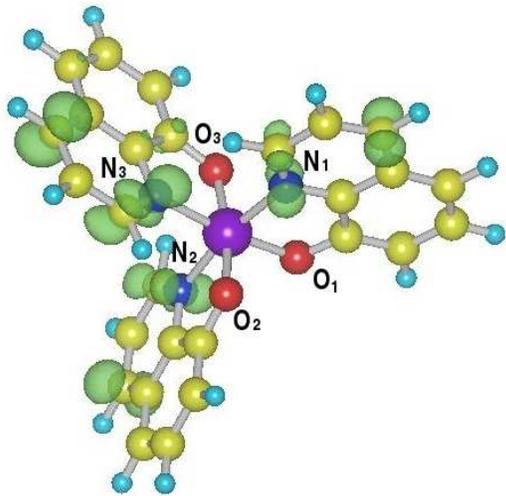}
\caption{(Color online) Geometry and charge-induced magnetization density of meridianal Alq$_{3}$. Al, O, N, C, and H atoms are shown as magenta, red, blue, yellow, and turquoise spheres, respectively. The magnetization density of the charged Alq$_3$ molecule with one extra electron is shown by the green hyper-surface. }
\label{nstr}
\end{figure}

The molecular structure of the meridianal isomer is shown in Fig.\ \ref{nstr}. The Al atom is surrounded by three nearly planar nitrogen atoms as well as by three nearly planar oxygens, being the link to the quinolines that consists of a pyridine and a benzene ring. 
In our study we first adopted the experimental structure (see, e.g., \cite{halls01}) and performed a full geometric relaxation for the neutral ground state. 
Subsequently, we used this DFT-computed ground state geometry and added extra charges (indicated later on by $q$) on the molecule to simulate the charge injection and repeated the full atomic relaxation procedure, assuming an equivalent neutralizing background in the simulation cell. 
Note that the added charge is not set to have any fixed spin.
Since in the charged molecule there will always be a net electric dipole moment, the dipole correction has also been included in our calculation. We have also tried positive charging, i.e., extracting  electron charge from the molecule. However, we find that in this case the molecule become very unstable and we are not  able to converge to a stable molecular geometry. This finding is in fact consistent with an experimental investigation that shows a very high positive charge injection energy as compared to a low energy for negative charge injection \cite{anazawa05}. 

%\section{Result and Discussions }

To start with, we compare our geometry optimized ground state structure of Alq$_3$ with available theoretical and experimental data, which are summarized in Table \ref{table1}. 
Previous studies showed that structural parameters of meridianal Alq$_3$ in good agreement with experiment could be obtained using Hartree-Fock calculations with the 3-21+G$^{**}$ basis set \cite{halls01}. A somewhat less good agreement was achieved with hybrid-functional B3LYP calculations \cite{johansson99,mart00}. 
It can be observed from Table \ref{table1} that our GGA calculations yield structural results which are in very good agreement with previous Hartree-Fock \cite{halls01} and recent  B3LYP calculations  \cite{mart00}. 
 
%%%%%

One observes from the structural data that the Al is bonded more closely to the oxygens than the nitrogens, and that it has different bond lengths to the six ligands. 
The N$_3$ atom has the largest separation from Al whereas N$_1$ and N$_2$ have similar bond lengths with Al. The three Al$-$O bond lengths are more equal. 
%Given this situation, 
Given the different Al$-$N bond lengths in an otherwise symmetric arrangement, an influence of charge doping on the molecule is not surprising. It is therefore interesting to examine the behavior of the molecule upon electron charging. 
%%%%%%%%%%%%%%%%%%%%%%%%%%%%%

\begin{table}[!tb]
\caption{\label{table1} Ground-state  Al$-$N and Al$-$O bond lengths (in {\AA}), for neutral meridianal Alq$_{3}$ from different theoretical calculations \cite{halls01,mart00,johansson99}, including this work, and experimental results \cite{halls01} (averaged over several available data sets).}
\begin{ruledtabular}
\begin{tabular}{cccccc}
  &HF/ & B3LYP & B3LYP & Exp. & GGA \\
% bond & 3-21+G**\footnotemark[1] & 6-31G(d)\footnotemark[2] &DNP\footnotemark[3]& ave.\footnotemark[1] & our cal. \\
 bond & 3-21+G**\cite{halls01} & 6-31G(d)\cite{mart00} &DNP\cite{johansson99}& ave.\cite{halls01} & our cal. \\
\hline
Al$-$N$_1$ & 2.063 & 2.08 & 2.019 & 2.044 & 2.063 \\
Al$-$N$_2$ & 2.033 & 2.06 & 2.019 & 2.025 & 2.050 \\ 
Al$-$N$_3$ & 2.117 & 2.13 & 2.051 & 2.079 & 2.103 \\
Al$-$O$_1$ & 1.826 & 1.86 & 1.839 & 1.847 & 1.865 \\ 
Al$-$O$_2$ & 1.866 & 1.89 & 1.868 & 1.867 & 1.888 \\
Al$-$O$_3$ & 1.856 & 1.88 & 1.864 & 1.856 & 1.886 \\
\end{tabular}
\end{ruledtabular}
\end{table}

 \begin{table}[!tb]
\caption{\label{table2}Ground state Al$-$N and Al$-$O bond lengths for meridianal Alq$_{3}$ upon injection with 0.5 and 1.0 electron, respectively.}
\begin{ruledtabular}
\begin{tabular}{ccc}
bond &  0.5 e (in \AA) & 1.0 e (in \AA) \\
\hline
Al$-$N$_1$ & 2.076 & 2.073 \\
Al$-$N$_2$ & 2.042 & 2.031 \\
Al$-$N$_3$ & 2.065 & 2.048 \\
Al$-$O$_1$ & 1.877 & 1.887 \\
Al$-$O$_2$ & 1.899 & 1.905 \\
Al$-$O$_3$ & 1.889 & 1.896 \\
\end{tabular}
\end{ruledtabular}
\end{table}

Table \ref{table2} shows the structural change of the molecule due to electron charging. Compared to the charge-neutral structure, the Al$-$N$_2$, Al$-$N$_3$ bond lengths are reduced, Al$-$N$_1$ becomes slightly longer. The Al$-$O$_1$ bond length also becomes longer, but the Al$-$O$_2$ and Al$-$O$_3$ distances are not changed as much. The change is gradual with the amount of charging of the molecule. 
%One interest observation is that 
The influence of the charge is most prominent for the N$_3$ atom having initially the maximum bond length. %The Al-O bond lengths become more 
 %are not affected appreciably and there is almost no asymmetry in the bond-lengths.
We further observe that the asymmetry of the three Al$-$O bond lengths is reduced with charging.

\begin{figure}[!tb]
  \includegraphics[width=0.9\linewidth]{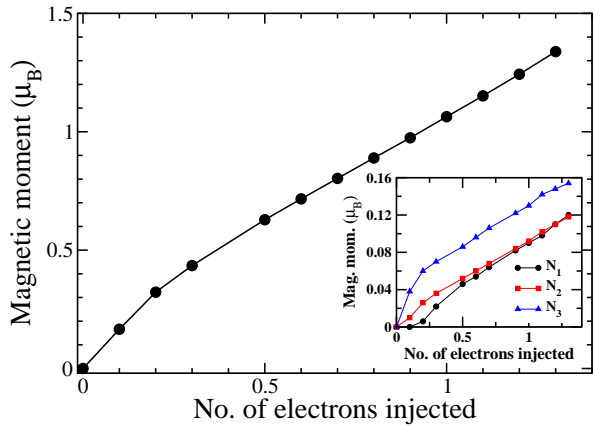}
\caption{\label{magmom} (Color online) Magnetic spin moment of Alq$_{3}$ as a function of injected electron charge. In the inset, the projected moments on the thee N atoms are shown with the added electron charge.}
\end{figure}

Our calculations reveal that upon electron charging the non-magnetic Alq$_3$ molecule spontaneously spin-polarizes.
Fig.\ \ref{magmom} shows the development of the molecule's spin moment due to the added charge. 
%The effect of extra charge on the molecule on the magnetic moment is shown in Fig.~\ref{magmom}. The charge neutral molecule is non-magnetic. For charged molecules, a finite magnetic moment is observed. 
The  magnetic moment of the molecule increases approximately linearly with the extra charge. This is a surprising result, because the extra charge could be equally distributed over the two spin channels of a molecular orbital, leading to a non-magnetic charged molecule, but this does not happen.
A key to understanding why this is the case is the asymmetric bonding in the neutral molecule. 
The added electron charge does not distribute equally over the three nitrogens, but it goes dominantly to the N$_3$ atom. In Fig.\ \ref{nstr} the magnetization density corresponding to $q=1.0$\,e is shown by the hyper-surface. It is clearly observed that the magnetization density resides prominently on the three N atoms. For this relatively high $q$ value, the carbon atoms of the pyridine ring display also some magnetization density, corresponding to magnetic moments of
% on the C atoms are 
0.08 $\mu_{B}$ on the average.
This magnetization density resides mainly on one of the carbon atoms nearest to nitrogen, as well as on the 3$^{rd}$ nearest neighbor carbon atom.
Neither O nor Al atoms carry a magnetic moment. 
%is not distributed equally in two spin channels. The asymmetry in the charge distribution results in a finite magnetic moment.  The magnetization density corresponding to q=1.0 is shown in Fig.~\ref{nstr}. It is clearly observed that the magnetization density resides prominently on the three N atoms connected to Al. O and Al atoms do not carry magnetic moments. The magnetic moments on the C atoms are 0.04 $\mu_{B}$ on the average.

From the above we deduce that the magnetic moment arising due to electron charging is related mostly to the N atoms. To understand this further, we have plotted the local magnetic moments on the three nitrogens as a function of extra electron charge on the molecule. This is shown in the inset of Fig.\ \ref{magmom}. The magnetic moments on the N$_1$ and N$_2$ atoms are varying in a similar manner  whereas the curve for N$_3$ is separated from that of N$_1$ and N$_2$. Over  the full range of charging, the N$_3$ atom has acquired a larger magnetic moment than the other two N atoms. This behavior corroborates with the structural change upon charging. As discussed earlier, the decrease of the Al$-$N$_3$ bond length is more significant than that of Al$-$N$_1$ and Al$-$N$_2$. This larger bond length reduction
appears to be related to the steeper increase in the magnetic moment of N$_3$. The Al$-$O bond lengths are not modified  appreciably with extra charge and consistently no spin-polarization on the oxygens is observed.

\begin{figure}[!tb]
   \includegraphics[width=0.9\linewidth]{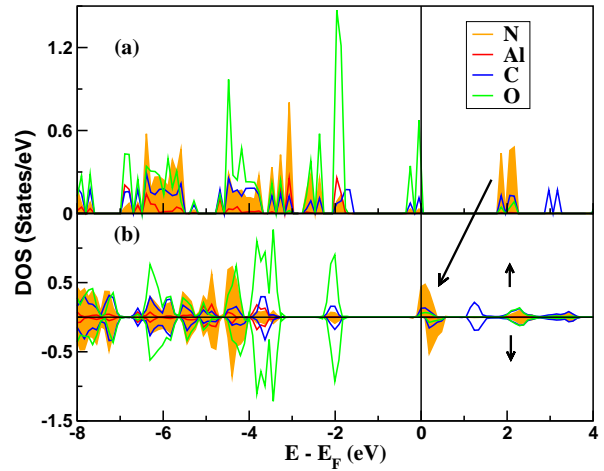}
\caption{\label{pdos1} (Color online) Atom projected density of states (DOS) of the Alq$_3$ molecule shown in (a) for the non-magnetic, charge neutral, and in (b) for the molecule with 1.0 electron doping. In (b), the spin-resolved partial DOS is shown. The arrow indicates the shift of the LUMO to the Fermi energy ($E_F$). Note that the N states of the LUMO become spin-polarized upon charging.}
\end{figure}

The atom-projected density of states (DOS) are shown in Fig.\ \ref{pdos1} for both neutral and a molecule charged with $q=1.0$\,e. The highest Al $p$ states occur at 2 eV binding energy. The Al orbitals are hence saturated with covalent bonds to the oxygens and nitrogens. The Al $3p$ shell is in fact over-saturated in the six-fold coordination, which explains why one nitrogen, here N$_3$, has a longer distance to Al. The quinoline ring is however structurally very stable, giving not much spacial freedom for a longer Al$-$N distance.
%We have also investigated removal of electrons from the molecule, however, we find that this makes the molecule {\it unstable}. 
As can be seen from Fig.\ \ref{pdos1}a for the neutral molecule, the HOMO level has mainly O-$p$ character whereas the LUMO consists of mainly N-$p$ states with a small admixture of O and C states. Al has a negligible contribution to both HOMO and LUMO.    
The calculated HOMO-LUMO gap is 1.87 eV. Upon charging, the N-$p$ states in the LUMO become spin-polarized and the HOMO-LUMO gaps for the spin-up and spin-down channels increase to 2.0 and 2.2 eV, respectively. The HOMO states remain practically non-polarized.
% the difference in the energy gaps in the two spin-channels occur.

The origin of the unexpected magnetization in the electron-injected state can be understood from the following observations. Due to the molecule's asymmetry the unoccupied N $p$-dominated states at 1.8 eV occur at slightly different energies, with the N$_3$ related states occurring at a somewhat lower energy.
Fig.\ \ref{pdos1} shows the unoccupied $p$ partial DOS of the atoms in the three quinoline rings. Note that we have broadened the DOS peaks, leading to a shift to lower energies. The LUMO consists only of atomic $p$-type orbitals, no other type is present. One can observe that electron doping commences with initially filling the empty $p_z$ levels of N$_3$. Hence, the moment on N$_3$ increases at first  steeply with doping. Due to the small asymmetry and the structural relaxation, the empty $p_x$ and $p_y$ states of N$_1$ and N$_2$ start to be filled more slowly. Note that these are equivalent to the N$_3$  $p_z$ orbitals by rotation. Electron doping of more than 0.5\,e leads to an increase of the moments on all three nitrogens at the same rate. For the narrow $p_z$ orbitals spin-polarized filling is more favorable than equal filling of both  spin-orbitals (Pauli principle). A somewhat larger electron filling would lead to an equal filling of both spin-orbitals on one nitrogen and thus to a non-magnetic state, but this does not happen, because the unoccupied $p_z$ orbitals of the other two nitrogens are preferably being filled first, which causes a small moment on N$_1$ and N$_2$ as well.  A ferromagnetic coupling between the three spin moments stabilizes finally a net spin-polarization on the whole molecule.

\begin{figure}[!tb]
   \includegraphics[width=0.9\linewidth]{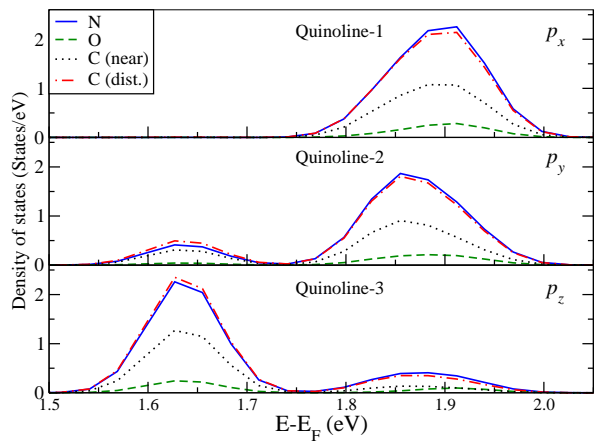}
\caption{\label{pdos2} (Color online) Atom-resolved $p$ partial density of states (DOS) of the lowest unoccupied molecular orbitals of Alq$_3$, shown separately for the three quinoline groups (labeled in Fig.\ \ref{nstr}). The  $p$ states of the quinoline-3 farthest away from Al have the lowest energy. 
}
\end{figure}

To conclude, our DFT calculations reveal that electron injection in Alq$_3$ leads to spontaneous spin-polarization. The predicted spin-polarization reaches to 1 $\mu_B$ per injected electron per molecule. 
The spin-polarization is found to originate from the preferential spin-polarized filling of empty N $p_z$ orbitals in the LUMO, in combination with a parallel coupling between the spins on the quinolines. 
%An asymmetry of the N $p_z$ orbital energies in the LUMO leads to an asymmetric distribution of spin-polarization on the nitrogens, with progressive electron injection the molecule becomes more symmetric.
This discovered induced spin-polarization of the normally non-magnetic Alq$_3$ molecule can be expected to be critically related to the extremely long spin relaxation times observed in Alq$_3$ \cite{pramanik07} as well as to be essential for  the observed large magnetoresistance \cite{xiong04}.  

We thank Y. Zhan and O. Eriksson and helpful discussions. This work has been supported financially by the Swedish Research Council (VR), SIDA, the C. Tryggers Foundation, and STINT. Computational support from the Swedish National Infrastructure for Computing (SNIC) is also acknowledged.

\end{document}